\documentclass[
prc,
preprintnumbers,amsmath,amssymb,floatfix,nofootinbib]{revtex4}

\usepackage{graphicx}
\usepackage{dcolumn}
\usepackage{bm}

\usepackage{amsmath}
\usepackage{psfrag}
\usepackage{color}
\usepackage{latexsym,amsfonts}
\usepackage{graphicx}

\providecommand{\href}[2]{#2}   

\def\b{b}
\def\a{a}
\def\x{\varkappa}
\def\ve{\varepsilon}
\def\mtx#1#2#3#4{\left(\begin{array}{cc}\!#1\!&\!#2\!\\\!#3\!&\!#4\!\end{array}\right)}

\begin{document}

\title{Two-photon exchange at low $Q^2$}
\author{Dmitry~Borisyuk}
\author{Alexander~Kobushkin}%
\email{kobushkin@bitp.kiev.ua}
\affiliation{Bogolyubov Institute for Theoretical Physics\\
Metrologicheskaya street 14-B, 03143, Kiev, Ukraine}

\date{\today}

\begin{abstract}
We study two-photon exchange for elastic electron-proton scattering
at low $Q^2$. Compact approximate formulae for the amplitudes are obtained.
Numerical calculations are done for $Q^2 \le 0.1 {\ \rm GeV}^2$
with several realistic form factor parameterizations, yielding similar results.
They indicate that the corrections to magnetic form factor can visibly
affect cross-section and proton radii. For low-$Q^2$ electron-neutron
scattering two-photon exchange corrections are shown to be negligibly small.
\end{abstract}

\maketitle

During last few years, two-photon exchange (TPE) in elastic electron-proton
scattering draw a lot of attention. The study of TPE was inspired
by the problem in the proton form factor (FF) 
measurements \cite{Arrington}. However, this is not the only field where TPE effects
can manifest oneself. The others include single-spin asymmetries \cite{SSA},
radiative corrections to parity-violating observables \cite{PV} and
determination of proton radii \cite{Rosen,Sick}. The latter is done via study
of elastic electron-proton scattering at low $Q^2$. The TPE corrections
are known to be important here, but the whole TPE amplitude
was never calculated in closed analytic form.

The TPE amplitude can be split into elastic (with proton intermediate state)
and inelastic contributions. Naturally, the elastic part should be dominant
at low energy.
In Ref.\cite{we} the elastic contribution was reduced to the twofold integral
containing FFs from the space-like region only. Nevertheless, the functions
under the integral are complicated and possess singularities, thus it is not
so easy to integrate numerically. At low energies, further simplification
is possible, yielding compact  analytic expressions for the TPE amplitudes.

We follow the notations of Ref.\cite{we}. The particle momenta are defined
according to $e(k) + p(p) \to e(k') + p(p')$. We also introduce
$P = (p+p')/2$, $K = (k+k')/2$ and $q = p'-p$. Scalar kinematic variables are
$\nu = 4PK = s-u$ and $t = q^2 = -Q^2$.

Neglecting the electron mass, the scattering amplitude 
can be written as
\begin{equation}
 {\cal M} = \frac{4\pi\alpha}{t} \bar u'\gamma_\mu u \ 
 \bar U' \left(\tilde F_1 \gamma^\mu - \frac{1}{4M} \tilde F_2
 [\gamma^\mu, \hat q] + \frac{1}{M^2} \tilde F_3 \hat K P^\mu \right) U.
\end{equation}
The invariant amplitudes $\tilde F_i$ are functions of $\nu$ and $t$.
In the Born approximation (BA) $\tilde F_1(\nu,t) = F_1(t)$,
$\tilde F_2(\nu,t) = F_2(t)$, $\tilde F_3(\nu,t) = 0$, where $F_{1,2}(t)$
are usual Dirac and Pauli proton FFs.
Alternatively, we may use any three independent linear combinations of
$\tilde F_i$; e.g. instead of $\tilde F_1$ and $\tilde F_2$ one may define
$\tilde G_M = \tilde F_1 + \tilde F_2$ and
$\tilde G_E = \tilde F_1 + \frac{t}{4M^2} \tilde F_2$, such that in BA
$\tilde G_E$ and $\tilde G_M$ are equal to electric and magnetic FFs.

The difference between $\tilde F_i$ and their Born values is proportional to
$\alpha \approx \frac{1}{137}$. Neglecting the terms of order $\alpha^2$ w.r.t.
the leading one, the cross-section for unpolarized particles is
\begin{equation} \label{xsect}
 d\sigma = 
 d\sigma_0 \left( \ve {\cal G}_E^2 - \tfrac{t}{4M^2} {\cal G}_M^2 \right),
\end{equation}
where 
\begin{eqnarray}
 {\cal G}_M = \tilde G_M + \ve \tfrac{\nu}{4M^2} \tilde F_3, &
 {\cal G}_E = \tilde G_E + \tfrac{\nu}{4M^2} \tilde F_3, &
\end{eqnarray}
and
\begin{equation}
 d\sigma_0 = \frac{8\pi\alpha^2}{1-\ve} \frac{4M^2}{(\nu-t)^2} \frac{dt}{t}.
\end{equation}
The invariant amplitudes are complex; in Eq.(\ref{xsect}) and everywhere below
their real parts are understood.

Eq.(\ref{xsect}) looks like Rosenbluth formula (certainly the experimental
separation of ${\cal G}_M$ and ${\cal G}_E$ is not possible, since both are
$\ve$-dependent beyond BA). Because of this simple form, we adopt the following
set of invariant amplitudes: ${\cal G}_E$, ${\cal G}_M$, and $\tilde G_M$.

In this paper we are interesting in TPE contributions
$\delta {\cal G} = (\delta {\cal G}_E,\delta {\cal G}_M,\delta \tilde G_M)$
at low $t$, $-t \ll 4M^2$. On the other hand, we neglect the electron
mass $m$, so it should be $4m^2 \ll -t$.

Low $t$ can be achieved in two ways: (i) at fixed energy and small scattering angle
and (ii) at fixed angle (or fixed $\ve$, which is nearly the same) and
low energy. We will consider the latter case. This implies
\begin{equation} \label{lownu}
 \nu = \x \sqrt{-t(4M^2-t)} \sim \sqrt{-t},
 {\rm\ where \ }
 \x = \sqrt{\frac{1+\ve}{1-\ve}}.
\end{equation}
%
In the Breit frame $\x = 1/\sin \frac{\theta}{2}$; this quantity was denoted
$x$ in Ref.\cite{Tomasi}.
Eq.(\ref{xsect}) suggests that at low $t$ one should primarily consider
the amplitude ${\cal G}_E$. However for generality we will study all three amplitudes.
The role of amplitude ${\cal G}_M$ increases for backward angles ($\ve \approx 0$)
and for neutron target ($G_E \approx 0$).
The amplitude $\tilde G_M$ may contribute to polarization observables.

\begin{figure}[t]
\centering
\psfrag{t1}{$t_1$}\psfrag{t2}{$t_2$}
\psfrag{t}{$t$}
\psfrag{I}{I}\psfrag{II}{II}\psfrag{III}{III}
\includegraphics[width=0.3\textwidth]{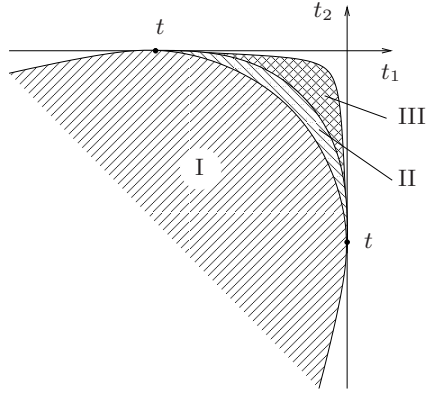}
\caption{The integration regions.}
\label{regions}
\end{figure}

The TPE contribution to any of the invariant amplitudes
can be written as \cite{we}
\begin{equation}
 \delta {\cal G} = 
  \frac{i\alpha}{2\pi^3} \sum_n \int
   \sum_{i,j=1}^2 A_{n,ij}(t_1,t_2) F_i(t_1) F_j(t_2) \frac{d^4p''}{t_1 t_2 D_n},
\end{equation}
where $n = 1,2,3,4,{\rm 4x}$;
\begin{eqnarray}
 & & D_1 = 1, \ \ D_2 = (P-p''+K)^2 - m^2,\ \ D_3 = p''^2-M^2,\\
 & & D_4 = [(P-p''+K)^2-m^2][p''^2-M^2],\ \ 
 D_{\rm 4x} = [(P-p''-K)^2-m^2][p''^2-M^2] \nonumber
\end{eqnarray}
and $t_1 = (p-p'')^2$, $t_2 = (p'-p'')^2$.
The factor $\frac{i\alpha}{2\pi^3}$ is introduced for convenience.
$A_{n,ij}$ are some polynomials in $t_1,t_2$, depending also on $\nu$ and $t$,
and originate from the numerators of TPE diagrams.
They are different for different amplitudes in the l.h.s., but
to simplify notation we do not attach an index to indicate this.
The crossing symmetry implies
\begin{equation}
 A_{1,2,3}(\nu) = - A_{1,2,3}(-\nu), \ \ \ A_{4}(\nu) = - A_{\rm 4x}(-\nu).
\end{equation}
Following Ref.\cite{we}, the integrals over $d^4p''$ can
be reduced to the twofold integrals over $t_1$, $t_2$:
\begin{equation} \label{main}
 \delta {\cal G} = 
  \frac{\alpha}{2\pi^3} \sum_n \intop_{t_1,t_2 \le 0} {\cal K}_n(t_1,t_2)
   \sum_{i,j=1}^2 A_{n,ij}(t_1,t_2) F_i(t_1) F_j(t_2) \frac{dt_1 dt_2}{t_1 t_2}, 
\end{equation}
where ${\cal K}_n$ are known functions, given in Eqs.(31-35) of Ref.\cite{we}.
It is convenient to define $A_{4\pm} = A_4 \pm A_{\rm 4x}$, 
${\cal K}_{4\pm} = \frac{1}{2} ({\cal K}_4 \pm {\cal K}_{\rm 4x})$, then
\begin{equation}
 A_4{\cal K}_4 + A_{\rm 4x}{\cal K}_{\rm 4x} = A_{4+}{\cal K}_{4+} + A_{4-}{\cal K}_{4-}.
\end{equation}
Now we will simplify all the terms of Eq.(\ref{main}) in the case of low $t$.
Due to $\theta$-functions, contained in ${\cal K}_n$, the integration in (\ref{main})
is done over three regions: (I) $x_\infty>0$, (II) $x_\infty<0<x_M$, $t_1+t_2-t>0$ and 
(III) $x_M<0<x_m$, $t_1+t_2-t>0$, see Fig.~\ref{regions}. In the regions II and III $t_{1,2} \sim t$
but in the region I $t_1$ and $t_2$ can be large. However the FFs entering
(\ref{main}) decrease rapidly as $t_{1,2} \to -\infty$, so the main contribution
in the region I also comes from $t_{1,2} \sim t$.
Assuming $t_{1,2} \sim t$, we obtain
\begin{eqnarray} \label{lowK}
\hskip-10mm&& {\cal K}_1 \sim 1,\ \ \ {\cal K}_2 \sim 1/t,\ \ \ {\cal K}_{4-} \sim 1/t\sqrt{-t},\\
\hskip-10mm&& {\cal K}_3 \approx \frac{-\pi^2 i}{4M\sqrt{-t}}
  [ \theta(x_\infty) + 2 \theta(-x_\infty)\theta(x_M)\theta(t_1\!+\!t_2\!-\!t) ]
  \approx \frac{-\pi^2 i }{4M\sqrt{-t}} \theta(x_\infty) \sim 1/\sqrt{-t}, \nonumber \\
\hskip-10mm&& {\cal K}_{4+} \approx \frac{-\pi^2 i}{2\sqrt{\!R}} {\rm sign}(t_1\!+\!t_2\!-\!t)
  [ \theta(x_\infty) + 2 \theta(-x_\infty)\theta(x_M)\theta(t_1\!+\!t_2\!-\!t) ]
  \approx \frac{-\pi^2 i}{2\sqrt{\!R}} \theta(x_\infty) {\rm sign}(t_1\!+\!t_2\!-\!t)
  \sim 1/t\sqrt{-t}, \nonumber
\end{eqnarray}
where $R \approx \nu^2\left(\tfrac{t_1+t_2-t}{2}\right)^2 - t_1 t_2 (\nu^2+4M^2t)$,
and the square root means arithmetic value.
Since the area of region II is small as $O(t)$, the term 
$\theta(-x_\infty)\theta(x_M)\theta(t_1\!+\!t_2\!-\!t)$ brings, after the integration,
an additional factor of $t$. Therefore we neglect it w.r.t.
$\theta(x_\infty)$. The ${\cal K}_n$ which are not written explicitly will
give negligible contribution, see below.

It is convenient to introduce new variables $\a$ and $\b$ such that
\begin{equation}
 t_1 = t\,(\a+1/4+\b), \ \ \ t_2 = t\,(\a+1/4-\b),
\end{equation}
then
\begin{equation}
 x_\infty = t\,(\a-\b^2), \ \ \ 
 dt_1 dt_2 = 2 t^2 d\a\, d\b,
\end{equation}
the condition $x_\infty \ge 0$ is equivalent to $\a \ge 0$, $\b^2\le\a$ and 
$t_{1,2} \sim t$ means $\a,\b \sim 1$.

Consider the quantities $A_{n,ij}$.
Taking into account (\ref{lownu}), the low-$t$ behaviour of $A_{n,ij}$,
for all three amplitudes, is
\begin{equation}
 A_1 \sim \sqrt{-t},\  A_2 \sim t\sqrt{-t},\  A_{4-} \sim t^2,\ 
 A_3 \sim \sqrt{-t},\  A_{4+} \sim t\sqrt{-t}.
\end{equation}
Comparing with (\ref{lowK}), we conclude that $A_1{\cal K}_1$, $A_2{\cal K}_2$
and $A_{4-}{\cal K}_{4-}$ give a contribution of order $O(\sqrt{-t})$ w.r.t 
$A_3{\cal K}_3$ and $A_{4+}{\cal K}_{4+}$, so further we will neglect
the former and consider the latter.
The explicit expressions for the leading terms in $A_{n,ij}$ are,\\
for the amplitude $\tilde G_M$:
\begin{equation}
 A_{3} = \frac{\nu(\a-1/4)}{\x^2-1} \mtx2110,\ \ \ 
 A_{4+} = \nu t \left[ \a + 1/4 - \frac{(\a-1/4)^2}{\x^2-1} \right] \mtx2110
 - \nu t \b \mtx01{-1}0,
\end{equation}
for the amplitude ${\cal G}_M$:
\begin{equation}
 A_{3} = \frac{\nu}{\x^2+1} \mtx2110,\ \ \ 
 A_{4+} = \frac{\nu t}{\x^2+1} \left[ \frac{2\x^2+1}{4} + \b^2 \right] \mtx2110
 - \nu t \b \mtx01{-1}0
\end{equation}
and for ${\cal G}_E$:
\begin{equation}
 A_{3} = \frac{\nu}{\x^2-1} \mtx1000,\ \ \ 
 A_{4+} = \nu t \left[ 1 - \frac{\a-1/4}{\x^2-1} \right] \mtx1000,
\end{equation}
where we use matrix notation
$A_{n} = \mtx{A_{n,11}}{A_{n,12}}{A_{n,21}}{A_{n,22}}$
for compactness.

Now let us consider the FFs
 $F_i(t_1)F_j(t_2) \equiv F_i(t\a\!+\!t/4\!+\!t\b) F_j(t\a\!+\!t/4\!-\!t\b)$.
For $\a \gg 1$ we have $\b \le \sqrt{\a} \ll \a$, so 
$F_i(t_1)F_j(t_2) \approx F_i(t\a+t/4)F_j(t\a+t/4)$.
In the main region of interest $\a \sim 1$ we expand $F$ in Taylor series
around $t\a+t/4$ and obtain
\begin{eqnarray}
 & & F_i(t_1)F_j(t_2) + F_i(t_2)F_j(t_1) \approx 2 F_i F_j + O\left(t/t_0\right)^2, \\
 & & F_i(t_1)F_j(t_2) - F_i(t_2)F_j(t_1) \approx t\b (F_i'F_j - F_j'F_i) + 
  O\left(t/t_0\right)^3, \label{ffasym}
\end{eqnarray}
 where FFs and their derivatives in the r.h.s are taken at $t\a+t/4$, and
$t_0$ is characteristic scale for FFs,
\begin{equation}
t_0 = 6/\langle r^2 \rangle \approx 0.3 {\rm\ GeV}^2 \text{ for proton.}
\end{equation}
Note that $t_0 \ll 4M^2$ (this is because proton radius is related 
to $pion$ rather than proton mass), and thus an additional requirement arises:
$(t/t_0)^2 \ll 1$.

The r.h.s. of Eq.(\ref{ffasym}) is of order $t/t_0$. In the presence of FF
scaling, $F_2/F_1 \approx {\rm const}$, it becomes much smaller, since
$F_i'F_j - F_j'F_i = F_i F_j (\ln F_i/F_j)'$, so we will assume
(\ref{ffasym}) to be negligible and thus put everywhere
\begin{equation}
 F_i(t_1)F_j(t_2) = F_i(t\a+t/4) F_j(t\a+t/4). 
\end{equation}
Replacing $F(t\a+t/4)$ by just $F(0)$ would be too rough an approximation,
since the FFs suppress the integrand at $\a\gg 1$.

After all the approximation made the FFs do not depend on $\b$ and the
integration over $\b$ can be done analytically:
\begin{eqnarray}
 J_3 & = & -\frac{2Mt^2\sqrt{-t}}{\pi^2 i} \int \frac{d\b}{t_1 t_2} {\cal K}_3 =
   \frac{1}{\a+1/4} \ln \left|\frac{\sqrt{\a}+1/2}{\sqrt{\a}-1/2}\right|, \\
 J_4 & = & - \frac{2Mt^3\sqrt{-t}}{\pi^2} \ {\rm Im} \int \frac{d\b}{t_1 t_2} {\cal K}_{4+} =
   \frac{1}{2\x} \, \frac{1}{\a^2-1/16}
     \ln \left|\frac{\a+1/4+\x\sqrt{\a}}{\a+1/4-\x\sqrt{\a}}\right|, \\
 J_4' & = & -\frac{2Mt\sqrt{-t}}{\pi^2} \ {\rm Im} \int d\b \, {\cal K}_{4+} =
   \frac{1}{2\sqrt{\x^2-1}} \ln \left|
     \frac{\a-1/4+\sqrt{\x^2-1}\sqrt{\a}}{\a-1/4-\sqrt{\x^2-1}\sqrt{\a}}\right|. 
\end{eqnarray}
Collecting all the expressions together, we obtain finally
\begin{eqnarray} \label{res}
 \delta \tilde G_M & = &
  \frac{2\alpha}{\pi} \x \int_0^\infty \left\{
   \frac{\a-1/4}{\x^2-1} J_3 + \left[\a+1/4 - \frac{(\a-1/4)^2}{\x^2-1} \right] J_4
   \right\} F_1 (F_1 + F_2) d\a, \\
 \delta {\cal G}_M & = & 
  \frac{2\alpha}{\pi} \x \int_0^\infty \left\{ \frac{1}{2} J_4 + 
    \frac{1}{\x^2+1} \left[ J_3 + (\a-1/4)(\a+3/4) J_4 - J_4' \right]
   \right\} F_1 (F_1 + F_2) d\a, \\ \label{resend}
 \delta {\cal G}_E & = & 
  \frac{\alpha}{\pi} \x \int_0^\infty \left\{
   \frac{1}{\x^2-1} J_3 + \left( 1 - \frac{\a-1/4}{\x^2-1} \right) J_4
   \right\} F_1^2 d\a,
\end{eqnarray}
where $F_{1,2} \equiv F_{1,2}(t\a+t/4)$.

The approximate formulae (\ref{res}-\ref{resend}) should work well for
\begin{equation}
 -t/4m^2 \gg 1,\ \ \ \sqrt{-t}/2M \ll 1,\ \ \ (t/t_0)^2 \ll 1.
\end{equation}
The infrared problems would arise at $t_1 = 0$, $t_2=t$ or $t_1 = t$, $t_2=0$,
that is, at $\a = 1/4$. However the singularity in (\ref{res}-\ref{resend}) at $\a = 1/4$ is
integrable, so the IR divergent terms do not appear. Actually such terms have
the form $\ln \frac{\nu-t}{\nu+t} \ln \lambda^2 \approx \frac{t}{\nu} \ln \lambda^2$
and are negligible in our approximation.

Consider the limit $t=0$. In this case $F_1 = 1$, $F_2 = \mu-1$, where $\mu$ is
magnetic moment. In other words, the result will be the same as for
point-like particle.
%
%
Taking into account that $\x > 0$, we perform the integration and obtain for $t=0$
\begin{equation} \label{res0}
 \delta \tilde G_M = \mu\alpha\pi \frac{\x+1/2}{\x+1},\ \ \
 \delta {\cal G}_M = \mu\alpha\pi \frac{\x+1/2}{\x^2+1},\ \ \
 \delta {\cal G}_E = \alpha\pi \frac{1/2}{\x+1}.
\end{equation}
%

The last result for $\delta {\cal G}_E$ corresponds to well-known second-order
correction to Coulomb scattering \cite{McKinley}.

One point is to be clarified here. The TPE contributions should be odd functions
of $\x$ \cite{Tomasi}. From this the authors of Ref.\cite{Tomasi} make
a (erroneous) conclusion, that TPE amplitudes should have the form
\begin{equation} \label{series}
 \delta {\cal G} = \x \sum_{k=0}^\infty c_k(t) \x^{2k},
\end{equation}
which is apparently not the case for (\ref{res0}).
The source of error is that the series (\ref{series}) converges only in
certain circle around $\x = 0$ in the complex $\x$ plane. The radius
of the circle is the distance to the nearest singularity of $\delta {\cal G}(\x)$.
Such singularities are branching points at $s=(m+M)^2$ and at $u=(m+M)^2$,
that is, at $\nu = s-u = \pm(4Mm+t)$; neglecting $m$, this yields
$ \pm \x = \sqrt{\frac{-t}{4M^2-t}} < 1 $. But the physical values are
$\x \ge 1$, see Eq.(\ref{lownu}). They therefore lie outside the convergence
circle, and Eq.(\ref{series}) is not valid for these $\x$.
\begin{figure}[t]
\centering
\includegraphics[width=0.45\textwidth]{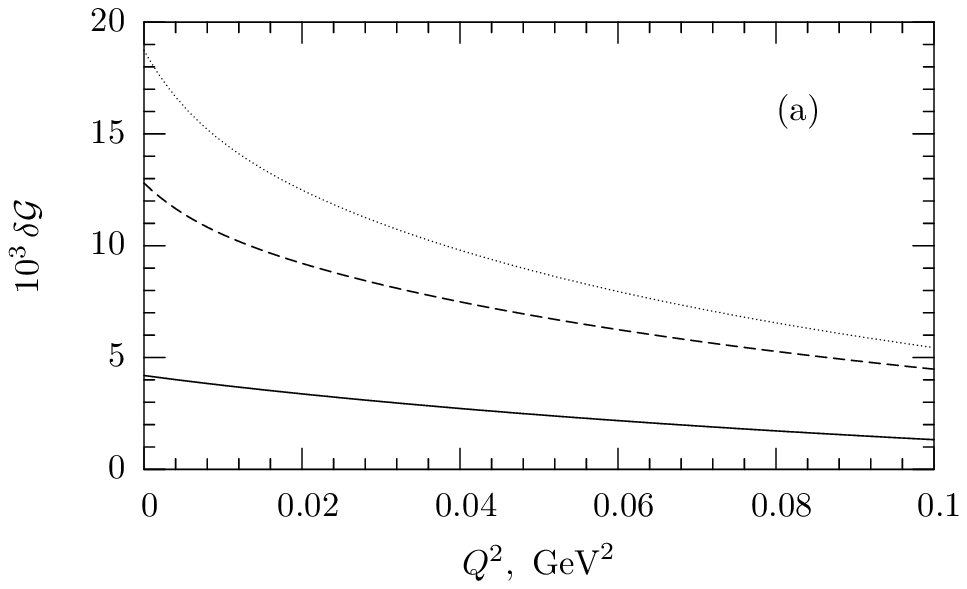}
\includegraphics[width=0.45\textwidth]{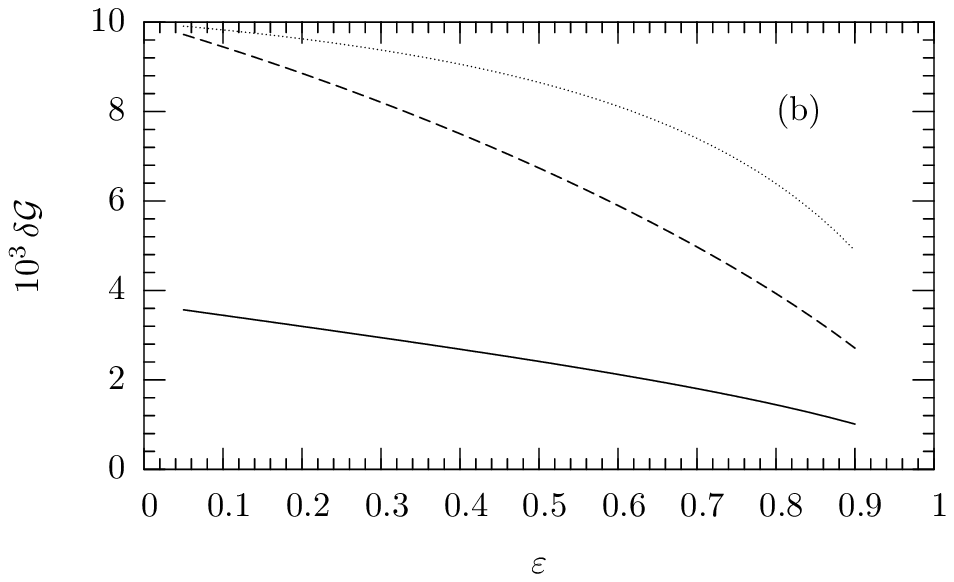}
\caption{TPE amplitudes at $\ve = 0.5$ (a) and at $Q^2 = 0.05 \ {\rm GeV}^2$ (b); $\delta {\cal G}_E$ (solid), $\delta {\cal G}_M/\mu$ (dashed) and $\delta \tilde G_M/\mu$ (dotted).}
\label{TPE}
\end{figure}
\begin{figure}[t]
\centering
\includegraphics[width=0.45\textwidth]{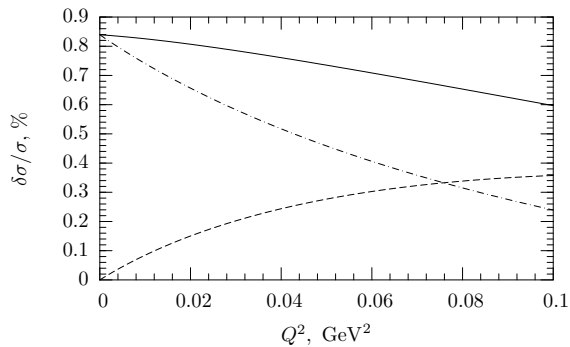}
\caption{TPE corrections to the cross-section at $\ve = 0.5$, electric (dash-dotted), magnetic (dashed) and total (solid).}
\label{dsig}
\end{figure}

The results of numerical evaluation of TPE amplitudes according to
Eqs.(\ref{res}-\ref{resend}) are shown in Fig.~\ref{TPE} (the amplitudes $\delta {\cal G}_M$ and $\delta \tilde G_M$ are divided by $\mu=2.79$). The calculation is done
with the dipole FFs. To check the sensitivity to the FF parameterization,
we repeated the calculation using FF fits \cite{Fits}; the resulting changes
in $\delta {\cal G}$ were small.

The TPE amplitudes are falling rapidly near $t=0$;
especially this pertains to the ``magnetic'' corrections $\delta {\cal G}_M$ and $\delta \tilde G_M$. Since the determination of rms radius involves FF derivative rather than FF itself, our results suggest that magnetic TPE correction
may be important here.
Roughly speaking, TPE corrections are to be subtracted from the experimentally 
measured FFs, resulting in slower decrease of ``corrected'' FFs. So we expect that inclusion of TPE should {\it lessen} proton rms radii.
Surprisingly, Refs.\cite{Rosen,Sick}, where TPE effects were calculated merely numerically, claim the increase of charge radius, by (0.008 - 0.013) fm \cite{Rosen} and by 0.0015 fm \cite{Sick}.

In Fig.~\ref{dsig} we show the TPE corrections to the cross-section at $\ve = 0.5$,
electric (coming from $\delta {\cal G}_E$), magnetic (coming from $\delta {\cal G}_M$)
and total. We see that magnetic contribution is of the same order as
the electric one and represents an essential piece of the total correction.
At lower $\ve$ it becomes even more important. 

The formalism developed here can be applied to electron-neutron scattering
as well.
In this case $G_E,F_1 \sim t$, and we see from (\ref{res}-\ref{resend})
that $\delta {\cal G}_M \sim t$ and $\delta {\cal G}_E \sim t^2$.
In other words, the TPE amplitudes are suppressed, with respect to the Born
amplitude, by a factor of $\alpha t$, which is much smaller than just $\alpha$.
Consequently, TPE corrections to the elastic $en$ scattering are negligibly
small at low $t$.

In summary, we obtained simple approximate formulae for TPE amplitudes of elastic electron-proton scattering at low $Q^2$. Numerical calculations indicate that the magnetic TPE corrections are relatively large and have a sharp $Q$-dependence, which can affect the determination of proton rms radii.

\end{document}